\documentclass[longbibliography,prb,twocolumn,showpacs,superscriptaddress,aps]{revtex4-1}
\usepackage[english]{babel}
\usepackage[dvips]{graphicx}
\usepackage{color}
\usepackage{latexsym}
\usepackage{amsmath}
\usepackage{amsthm}
\usepackage{amsfonts}
\usepackage{amssymb}
\usepackage{romannum}
\usepackage{mathbbol}
\usepackage[colorlinks=true,citecolor=blue,linkcolor=blue]{hyperref}

\newcommand{\eps}{\varepsilon}
\newcommand{\Remove}[1]{}

\newcommand{\bs}[1]{\boldsymbol{#1}}

\begin{document}

\pagenumbering{arabic}

\title{Transport properties of vertical heterostructures under light irradiation}
%
%
\author{P. Stadler}
\affiliation{Department of Microtechnology and Nanoscience (MC2), \\ Chalmers University of Technology, S-412 96 G{\"o}teborg, Sweden}

%
\author{M. Fogelstr{\"o}m}
\affiliation{Department of Microtechnology and Nanoscience (MC2), \\ Chalmers University of Technology, S-412 96 G{\"o}teborg, Sweden}
\begin{abstract}
Electronic and transport properties of bilayer heterostructure under light irradiation are of fundamental interest to improve functionality of optoelectronic devices. We theoretically study the modification of transport properties of bilayer graphene and bilayer heterostructures under a time-periodic external light field. The bulk electronic and transport properties are studied in a Landauer-type configuration by using the nonequilibrium Green's function formalism. To illustrate the behavior of the differential conductance of a bilayer contact under light illumination, we consider tight-binding models of bilayer graphene and graphene/hexagonal boron-nitride heterostructures. The non-adiabatic driving induces sidebands of the original band structure and opening of gaps in the quasienergy spectrum. In transport properties, the gap openings are manifested in a suppression of the differential conductance. In addition to suppression, an external light field induces an enhancement of the differential conductance if photoexcited electrons tunnel into or out of a Van~Hove singularity.

\end{abstract}
\date{\today}
\maketitle
%
%
%
\section{Introduction}
\label{sec:intro}
Two-dimensional materials have attracted enormous attention due to their unique mechanical, optical and electronic properties~\cite{Ferrari:2015js}. Out of the great variety of two-dimensional materials, graphene is the most studied one owing to its high electronic mobility~\cite{bolotin:2008}, large tensile strength~\cite{lee:2008} and strong light-matter interaction~\cite{nair:2008}. 
After isolation of graphene, the attention gradually shifted towards other two-dimensional materials displaying diverse properties which complement those of graphene. For example, in comparison to gapless graphene  hexagonal boron-nitride is a wide-gap insulator~\cite{giovannetti:2007,cassabois:2016}. 

Besides the study of monolayer materials, the large range  of different two-dimensional materials enables the study of hybrid systems.  Experimental  techniques have been developed to stack two dimensional materials on top of one another in an atomically  precise sequence~\cite{wang:2013,geim:2013}. The resulting artificial materials are named van der Waals (vdW) heterostructures and are characterized by strong interatomic bonds in the two-dimensional layer and weak interlayer coupling~\cite{liu:2016,novoselov:2016}. The weak vdW interactions allow the integration of different two-dimensional materials without the constraints of lattice matching imposed by conventional heterostructures.

The possibility to assemble different two-dimensional materials in vertical heterostructures enables the combination of single layer properties into one device. Combined with the optical properties of two-dimensional materials~\cite{nair:2008}, it makes them an attractive building block for optoelectronic devices such as photodetectors~\cite{mueller:2010,withers:2013,Liu:2014ku}. At the core of enhanced functionality lies the interaction of light and matter. It is hence of fundamental interest to study the effects of light in bilayer heterostructures.

Within the field of two-dimensional materials, the interaction of light and matter has not only to be studied to enhance or engineer new functionality of optoelectronic devices but also to reveal physical properties in these materials. For example, it has been experimentally shown that band gaps open in the energy spectra by irradiating graphene with a time-periodic potential~\cite{calvo:2011,calvo:2012,calvo:2013,atteia:2017,Islam:2018,torres:2014}. It has also been suggested that the opening of the gap is accompanied by a phase transition from a topological trivial to a topological one~\cite{oka:2009,kitagawa:2010,lindner:2011}. 
The topological nature of edge states in nanoribbons and signatures in transport properties has been extensively studied in Refs.~\onlinecite{torres:2014,Islam:2018,atteia:2017,piskunow:2014,usaj:2014,piskunow:2015,tenenbaum:2013,sentef:2015,kundu:2014,yang:2016,fruchart:2016,kitagawa:2011,gu:2011} in the absence of dissipation. The effect of dissipation on the transport properties can be studied by coupling the electronic system to a bath of phonons similar to the approaches in Refs.~\onlinecite{dehghani:2014,dehghani:2015,dehghani:2015b,dehghani:2016,iadecola:2015,chen:2018}. 
Electronic and transport properties in bilayer graphene under light irradiaton have been discussed in Refs.~\onlinecite{abergel:2009,abergel:2011,lago:2017,suarez:2012, calvo:2021, luo:2021}. Instead of a time-periodic external light field, time-dependent modulation
of gate or contact potentials and the accompanied photo-assisted tunneling of electrons lead to a great variety of interesting phenomena\cite{korniyenko:2016_1,korniyenko:2016_2,hammer:2013,pedersen:1998,platero:2004,kohler:2005}.

\begin{figure}[b!]			
	\includegraphics[width=0.7\linewidth,angle=0.]{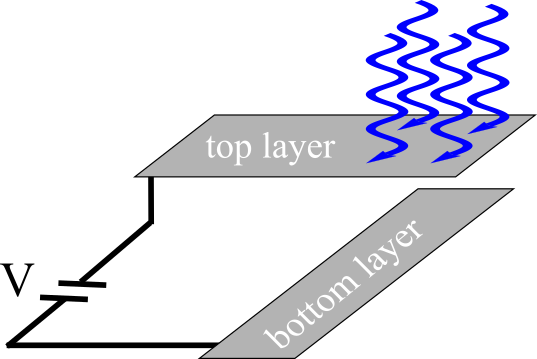} 
	\caption{Bilayer heterostructure under light illumination close to the overlap of the top and bottom layer. A voltage $V$ is applied between the two layers. }   
	\label{fig:fig1}
\end{figure}

Based on the growing interest in stacking 2D materials, we study transport properties in bilayer heterostructures under light irradiation as shown in Fig.~\ref{fig:fig1}. Previously, we considered a vertical heterostructure in a quantum-point contact configuration in which the top layer is irradiated with light~\cite{stadler:2020}. Compared to the study in Ref.~\onlinecite{stadler:2020}, the objective of this work is to consider a Landauer-type configuration consisting of a central region and reservoir leads. The leads remain in equilibrium if the central region is illuminated with light such the distribution functions in the leads are well-defined. As particular examples, we consider bilayer graphene and graphene/hBN heterostructures and discuss the modification of bulk transport properties under light irradiation. 

As the main result, we obtain that the differential conductance is suppressed or enhanced by interaction with the light. The suppression occurs for voltages corresponding to a gap opening in the quasienergy spectra of the bilayer heterostructure. The enhancement of the conductance is related to the tunneling into or out of a Van~Hove singularity which provides a large density of states. Although various aspects gap opening and the suppression of the differential conductance has been discussed in literature~\cite{calvo:2011,calvo:2012,calvo:2013,atteia:2017,Islam:2018,torres:2014}, here we extend previous results to bilayer graphene and bilayer heterostructures and discuss processes leading to an enhancement of the differential conductance.

The paper is structured as follows. In Sec.~\ref{sec:model}  and \ref{sec:NEGF}  we introduce
the model Hamiltonian and the nonequilibrium transport formalism which is used to calculate the differential conductance. Section~\ref{sec:results} contains the results of the paper: In Sec.~\ref{sec:resultsparameters}, we discuss a parameter regime which can be achieved by state-of-the-art experiments in bilayer heterostructure. Sections.~\ref{sec:resultsdos} and \ref{sec:resultscond} investigate in detail the signatures of the light-matter interaction in the differential conductance and explain the features using the momentum- and energy-dependent density of states. In Sec.~\ref{sec:conclusions}, we summarize and conclude.

%
%
\section{Model}
\label{sec:model}
\begin{figure}[b!]			
	\includegraphics[width=0.7\linewidth,angle=0.]{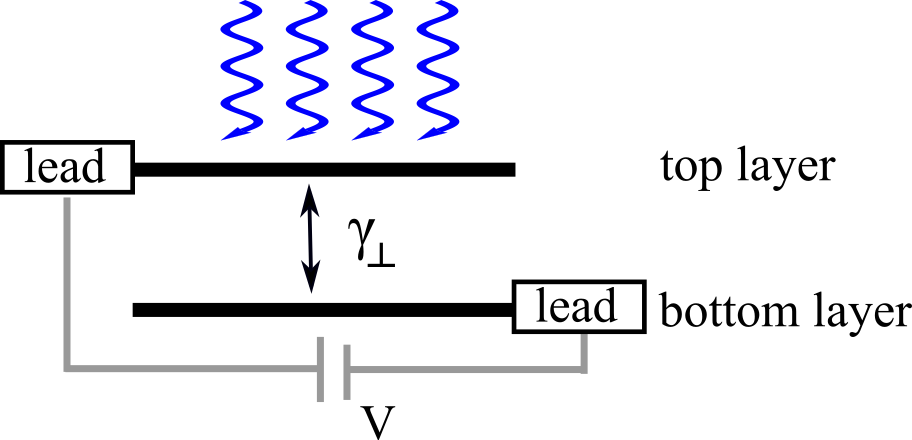} 
	\caption{Setup for the calculation of the differential conductance in a bilayer heterostructure consisting of a central region with a top and a bottom layer. The central region is illuminated with light and connected to leads which remain in equilibrium under light illumination. A voltage $V$ is applied between the top and bottom layer and the interlayer coupling is described by $\gamma_{\perp}$. }   
	\label{fig:fig2}
\end{figure}

We consider a bilayer heterostructure with a top and bottom layer connected to a top and bottom lead as shown in Fig.~\ref{fig:fig2}. An external electric field is applied perpendicular to the central region consisting of the bilayer heterostructure. The leads serve as reservoir leads and are assumed to remain in equilibrium if the central region is irradiated with light. 
%
The Hamiltonian can be composed of leads, a central region ($c$) and tunneling between the regions as 
\begin{equation}
\hat{H}(t) =  \hat{H}^{}_{\text{leads}}+ \hat{H}_{c}(t) +  \hat{H}_{\text{tun}} \, .
\label{eq:H}
\end{equation}
%
The Hamiltonian of the central region can be further divided into a top ($T$) and bottom ($B$) layer under light irradiation ($I$) and tunneling between the layers, 
\begin{equation}
\label{eq:Hct}
\hat{H}_{c}(t) =  \hat{H}^{T}_0 + \hat{H}^{B}_0 +\hat{H}_{\text{tun}}^{TB} + \hat{H}^{}_I(t)\, .
\end{equation}
$\hat{H}_0^{T,B}$ describes the bare top and bottom layers which we model as graphene and hexagonal boron-nitride. In the sublattice basis of the bilayer heterostructure, the Hamiltonian can be written as 
\begin{multline}
\hat{H}_0^{\alpha}=\sum_{\bs{k}} 
\eps^{a}_{\alpha}\hat{a}^{\dagger}_{\bs{k},\alpha} \hat{a}^{}_{\bs{k},\alpha} +
\eps^{b}_{\alpha} \hat{b}^{\dagger}_{\bs{k},\alpha} \hat{b}^{}_{\bs{k},\alpha} 
\\
+\left[\gamma_{\parallel}f_{}^{}(\bs{k})  \hat{a}^{\dagger}_{\bs{k},\alpha} \hat{b}^{}_{\bs{k},\alpha} + \textrm{H.c.} \right]
\end{multline}
with $\alpha=(T,B)$, the in-plane overlap energy $\gamma_{\parallel}$ of nearest neighbor atoms, and the operators 
$
\hat{a}^{}_{\bs{k},\alpha}
$
and
$
\hat{b}^{}_{\bs{k},\alpha}
$
of sublattice $a$ and $b$, respectively.	
	
The energies of the electrons in the 2$p_z$ orbitals are denoted as $\eps^{a}_{\alpha}$ and
$\eps^{b}_{\alpha}$.
%
For next-nearest neighbor coupling, the factor $f(\bs{k})$ is given by
$
f(\bs{k}) = \sum_i \mathrm{e}^{i \bs{k} \bs{\delta}_i }
$
with the vectors $\bs{\delta}_i$ connecting neighboring carbon atoms. These vectors are given by $\bs{\delta}_1 = \frac{a}{2} ( 1 ,   \sqrt{3} )$, $\bs{\delta}_2= \frac{a}{2}(1 , \mathord-\sqrt{3} )$  and $\bs{\delta}_3 = ( -1 , 0 )$ and the nearest neighbor distance between atoms $a = 1.42 \textrm{ \AA}$.

\begin{figure}[b!]			
	\includegraphics[width=1\linewidth,angle=0.]{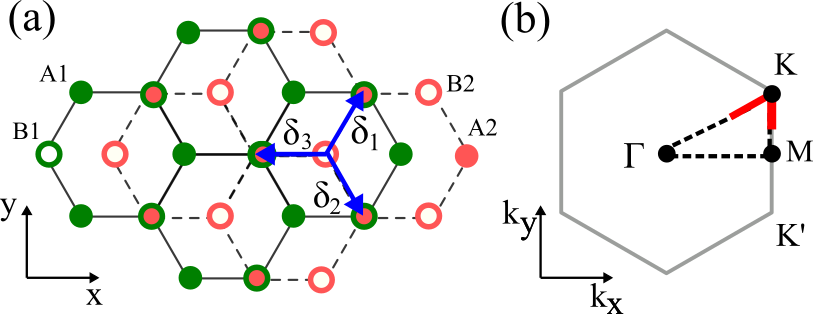} 
	\caption{(a) Bilayer heterostructure consisting of two hexagonal lattices. Triangular sublattices $A$ and $B$ are illustrated with full and empty circles. In the upper (lower) layer we write the sublattices as $A1$ ($A2$) and $B1$ ($B2$), respectively. The blue vectors $\bs{\delta}_i$ ($i=1,2,3$) connect the in-plane neighboring atoms. We consider both bilayer graphene and graphene/hBN heterostructures. (b) First Brillouin zone of bilayer heterostructure with the symmety points $\bs{\Gamma}$, $\bs{M}$, $\bs{K}$ and $\bs{K^{\prime}}$. The red path close to the $\bs{K}$-point indicated the momentum dependence of the density of states in Figs.~\ref{fig:fig4} and~\ref{fig:fig5}. }   
	\label{fig:fig3}
\end{figure}

The third term in Eq.~\eqref{eq:Hct} describes the coupling between the top and bottom layer. We consider a $AB$ stacked heterostructure in which an atom in the top layer of sublattice $a$ is over an atom of sublattice $b$ in the bottom layer and the atom of the top layer is over the hollow site of the lower layer [see Fig.~\ref{fig:fig3}(a)]. The Hamiltonian is 
\begin{equation}
\hat{H}_{\mathrm{tun}}^{TB} = \gamma_{\perp}\sum_{\bs{k}}
\hat{a}^{\dagger}_{\bs{k},T} \hat{b}^{}_{\bs{k},B} + \textrm{H.c.}
\end{equation}
with  out-of-plane coupling between the layers $\gamma_{\perp}$.

The last term in Eq.~\eqref{eq:Hct} describes the light-matter interaction of the bilayer heterostructure which we take into account by a minimal substitution $\hat{\bs{k}} \rightarrow \hat{\bs{k}} -e\bs{A}(t)$. In our study, we model the light classically in the dipole approximation since the momentum of the photon is negligible compared to the momentum of the electrons~\cite{Stroucken:2011,Malic:2011}. We are interested in the effects of circularly polarized light and take the vector potential to be 
$
\bs{A}(t)=(A_x\, \mbox{cos}(\omega t), A_y\, \mbox{sin}(\omega t), 0)
$
with clockwise polarization.  
The vector potential can be rewritten as
$
\bs{A}(t) = (A_0/2) (\bs{A}^{+} e^{i\omega t}+ \bs{A}^{-} e^{-i\omega t} )
$ 
proportional to positive $\bs{A}^{+} = (1,-i,0)$ and negative $\bs{A}^{-} = (1,i,0)$ photon frequencies $\omega$, respectively. 
In the sublattice basis, the light-matter interaction is
%
\begin{equation}
\label{eq:Hlightmatter}
\hat{H}^{}_I(t) = \sum_{\bs{k}, \alpha=T,B}  \gamma_{\alpha} M_{\bs{k},\alpha}^{} \hat{a}^{\dagger}_{\bs{k},\alpha} \hat{b}^{}_{\bs{k},\alpha} 
\end{equation}
with optical matrix elements separated into positive and negative frequencies as
$
M_{\bs{k},\alpha}^{}(t) = M_{\bs{k},\alpha}^{+}  e^{i\omega t} + M_{\bs{k},\alpha}^{-}  e^{-i\omega t}
$
and the coupling strength
$
\gamma_{\alpha} = (e/m)\hbar K_{\alpha} A_0/2
$
in units of energy. 
The elements 
$
M_{\bs{k},\alpha}^{\pm} 
$
are obtained from 
$
M_{\bs{k},\alpha}^{\pm} =  \bs{M}_{\bs{k}} \bs{A}^{\pm} 
$
with the in-plane vector
$
\bs{M}_{\bs{k},\alpha} =  \nabla_{\bs{k}} f(\bs{k})/a
$~\cite{Stroucken:2011,stadler:2020}.
The factor $K_{\alpha}$ is
$
K_{\alpha} = \int d\bs{x} \, \Phi_{\alpha}(\bs{x}) \partial_x \Phi_{\alpha}(\bs{x}-
\vert \bs{\delta}_i \vert \hat{\bs{e}}_x )
$ 
with the unit vector in x-direction $\hat{\bs{e}}_x$ and the
 $2p_z$-wave function $ \Phi_{\alpha}(\bs{x})$. In the following we assume that 
 $K_{T}=K_{B}$.

\section{Nonequilibrium transport formalism} 
\label{sec:NEGF}
In this paper, we study the transport properties of a bilayer heterostructure under light irradiation in a Landauer-type configuration where the central region is attached to leads as shown in Fig.~\ref{fig:fig2}. The irradiation is limited to the central region and dissipation of excess energy takes place in the leads allowing to apply a ballistic formulation of the transport problem.

The dc-current in a periodically driven systems can be written in terms of the Floquet Green's function 
$
\hat{{G}}^{n,R}_{\boldsymbol{k}}(\varepsilon)
$
which represent the Fourier transform of the retarded Green's function. To define the retarded Green's function, we introduce the operator in sublattice space of the top and bottom layer as
$
\hat{\psi}_{\boldsymbol{k}^{},\alpha}^{\dagger}
=
(
\hat{a}^{\dagger}_{\bs{k},\alpha} \,\, \hat{b}^{\dagger}_{\bs{k},\alpha}
)
$ ($\alpha=(T,B)$). 
The retarded Green's function is given by
\begin{equation}
\label{eq:GRt}
\check{G}^{R}_{\boldsymbol{k}}(t,t^{\prime}) = 
\begin{pmatrix} 
\hat{G}^{R}_{\boldsymbol{k},T,T}(t,t^{\prime}) 
&
\hat{G}^{R}_{\boldsymbol{k},T,B}(t,t^{\prime})
\\
\hat{G}^{R}_{\boldsymbol{k},B,T}(t,t^{\prime})
&
\hat{G}^{R}_{\boldsymbol{k},B,B}(t,t^{\prime})
\end{pmatrix}
\end{equation}
with 
$
\hat{G}^{R}_{\boldsymbol{k},\alpha,\alpha^{\prime}}(t,t^{\prime}) =
-i \theta(t-t^{\prime})\langle \{ \hat{\psi}^{}_{\boldsymbol{k},\alpha}(t) ,  \hat{\psi}_{\boldsymbol{k}^{},\alpha^\prime}^{\dagger}(t^{\prime}) \} \rangle  
$
and the anti-commutator $\{\dots\}$.
Since the Hamiltonian depends explicitly on time, we write the Floquet Green's functions as
$
\hat{{G}}^{R}_{\boldsymbol{k}}(t,t^{\prime}) = \sum_n \int \frac{d\varepsilon}{2\pi} \hat{{G}}^{n,R}_{\boldsymbol{k}}(\varepsilon) e^{-i \varepsilon \tau} e^{-i n \omega T}
$
with the relative time $\tau=t-t^{\prime}$, the center-of-mass time $T= (t+t^\prime)/2$~\cite{Rammer:2007}, and the index $n$ 
corresponding to the number of absorbed and emitted photons, respectively.
We then derive the Dyson equation~\cite{Cuevas-Scheer:2010,Rammer:2007} which after transformation to Fourier space is given by
\begin{align}
\hat{G}^{n,R}_{\bs{k}}(\eps) = \hat{g}^{n,R}_{\bs{k}}(\eps) 
+&
{\hat{g}^{n,R}_{\bs{k}}} (\eps) \hat{M}^{+}_{\bs{k}} \hat{G}^{n+1,R}_{\bs{k}}(\eps)
\nonumber 
\\ 
+&
{\hat{g}^{n,R}_{\bs{k}}}(\eps) \hat{M}^{-}_{\bs{k}} \hat{G}^{n-1,R}_{\bs{k}}(\eps) \, .
\label{eq:GR}
\end{align}
%
The matrices in Eq.~\eqref{eq:GR} have the same structure as the matrices in Eq.~\eqref{eq:GRt}. 
The unperturbed Green's function 
$
\hat{g}^{n,R}_{\bs{k}} (\eps)
$
is
\begin{equation}
\label{eq:GRrecursive}
\check{g}^{{n,R}}_{\bs{k}} (\eps)
=
\begin{pmatrix}
\eps^{n,a}_{T}  & -f(\bs{k})  & 0 & -\gamma_{\perp} 
\\
-f^*(\bs{k})  & \eps^{n,b}_{T} & 0 & 0 
\\
0 & 0 &  \eps^{n,a}_{B}  & -f(\bs{k}) 
\\  
-\gamma_{\perp}  & 0 &  -f^*(\bs{k})  & \eps^{n,b}_{B}  
\end{pmatrix}^{-1}
\end{equation}
with
$
\eps^{n,\beta}_{\alpha} = \eps + n\omega - \eps^{\beta}_{\alpha} 
$
and the index $\beta$ refering to sublattice $a$ or $b$, respectively. 
The matrices $ \hat{M}^{\pm}_{\bs{k}}$ in Eq.~\eqref{eq:GR} are
\begin{equation}
\hat{M}^{\pm}_{\bs{k},\alpha,\alpha}
= \gamma_{\alpha}
\begin{pmatrix}
0  & {M}^{\pm}_{\bs{k},\alpha} 
\\
{M}^{\pm}_{\bs{k},\alpha} & 0
\end{pmatrix}
\end{equation}
with the elements 
$
 {M}^{\pm}_{\bs{k},\alpha}
$
defined after Eq.~\eqref{eq:Hlightmatter} and 
$
{M}^{\pm}_{\bs{k},T,B} =
{M}^{\pm}_{\bs{k},B,T}
=0
$. 
Equation~\eqref{eq:GR} constitutes a recursive equation which can be solved by matrix inversion in the space of sidebands and sublattices, or an iterative technique~\cite{cuevas:2001,stadler:2020}.

The time-averaged current
$
I = (1/T) \int_0^T dt\, I(t)
$
between the leads can be expressed as~\cite{pedersen:1998,arrachea:2006}
\begin{equation}
\label{eq:Ib}
I = \frac{2e^2}{h}\sum_{n\bs{k}} \int d\eps \,  T^{n}_{\bs{k}}(\eps) (f_T(\eps)-f_B(\eps)) \, ,
\end{equation}
with the transmission 
$
T^{n}_{\bs{k}}(\eps) = {\Gamma}_T {\Gamma}_B \text{Tr} 
\vert \hat{\mathcal{G}}^{n,R}_{\boldsymbol{k},T,B}(\varepsilon) \vert^2
$.
Since the leads are non-irradiated, their occupations is well-defined by
the Fermi function of the top and bottom $f_{\alpha}(\eps)=\{1+\mathrm{exp}[(\eps-\mu_{\alpha})/k_B T]\}^{-1}$ with the chemical potential on the top and bottom layer $\mu_{\alpha}$ at temperature $T$.
The trace in the transmission is calculated in sublattice space and we have modeled the leads in the wide-band approximation.  In this approximation, the leads will induce a broadening of states in the central region which is defined by $\Gamma_{\alpha}=\rho(\eps_F) \vert t_{c\alpha} \vert^2$ with the density of states $\rho_{\alpha}(\eps_F)$ at the Fermi energy of lead $\alpha$ and the coupling $t_{c\alpha}$ between the leads and the top and bottom layer respectively. 
The assumptions of energy-independent coupling matrices and unitarity of the scattering matrix for incoming and outgoing waves leads to a vanishing time-averaged current in the absence of a voltage avoiding charge pumping effects~\cite{gu:2011,torres:2014}.

In this paper, we are interested in the bulk transport properties and study the differential conductance at zero temperature. Assuming the voltage to be applied on the top lead, the differential conductance is 
\begin{equation}
\label{eq:diffcond}
 \frac{dI}{dV}= \frac{2e^2}{h}\sum_{n\bs{k}} T_{\bs{k}}^{n}(eV)  \, \, ,
\end{equation}
with the transmission 
$
T_{\bs{k}}^{n}(\eps)
$
depending on the Green's function connecting the top and bottom layer.

\begin{figure*}[t!]			
	\includegraphics[width=0.47\linewidth,angle=0.]{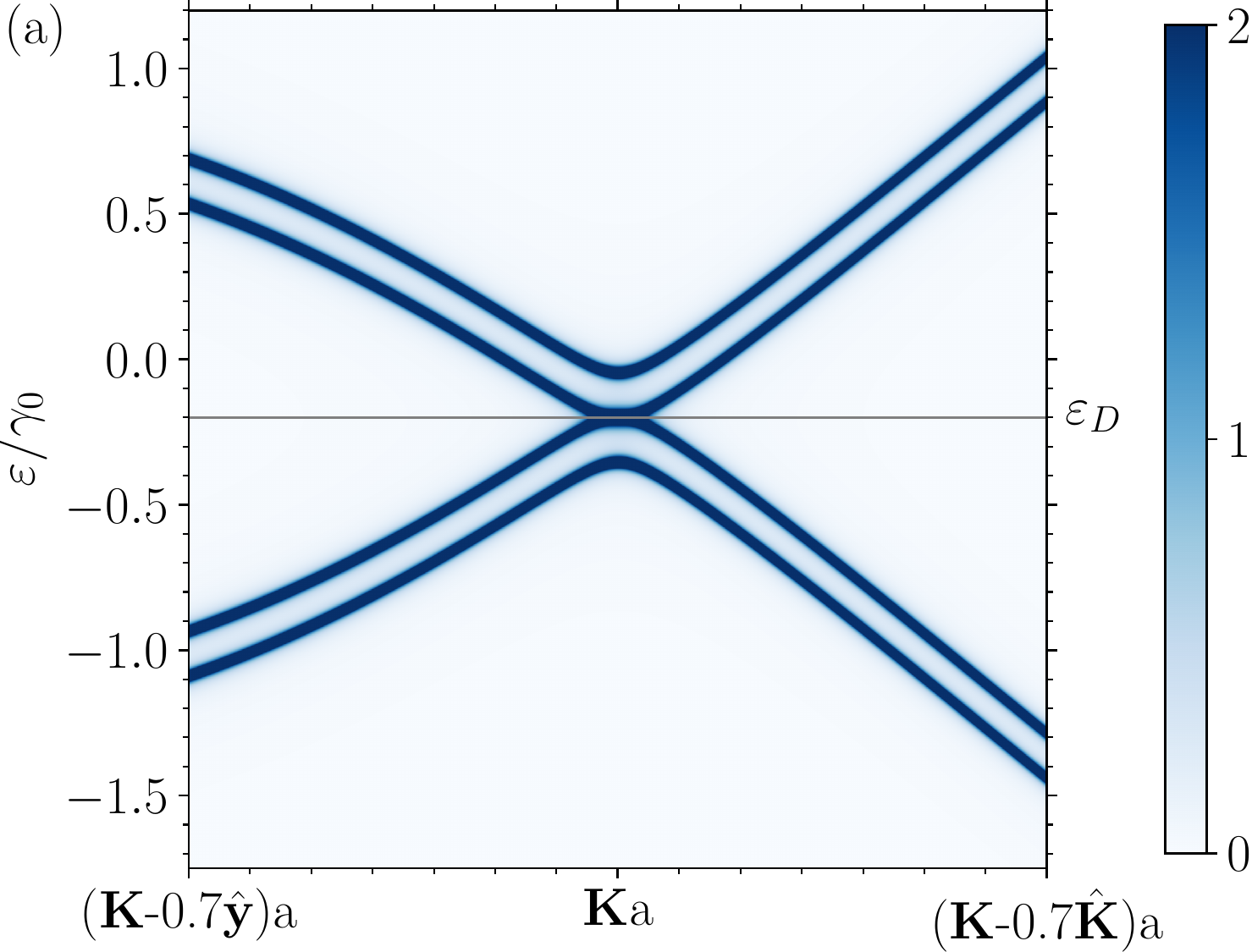} 
	\hspace{15pt}
	\raisebox{0mm}{\includegraphics[width=0.47\linewidth,angle=0.]{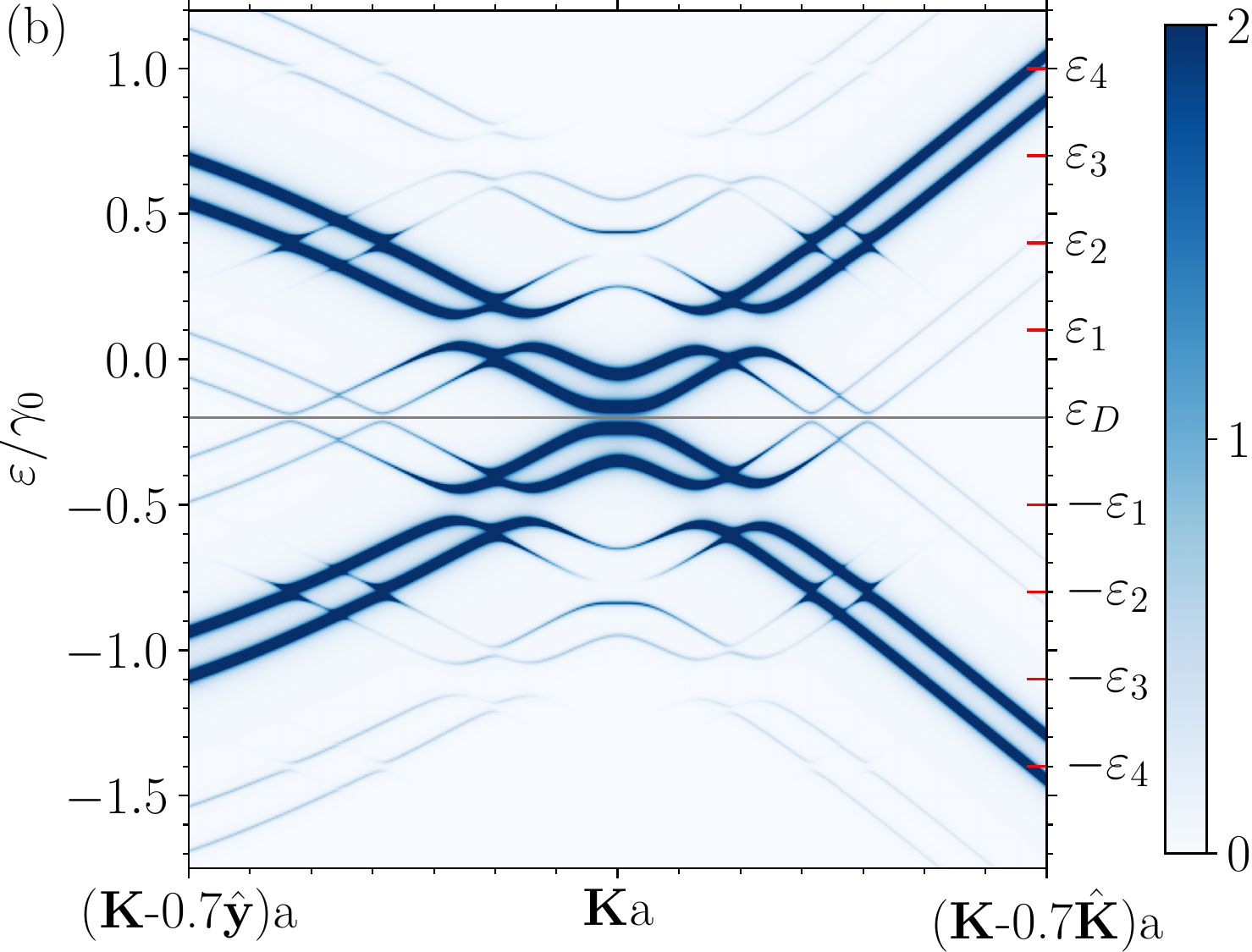}}
	\caption{Spectral density of states $\rho_{\bs{k}}(\eps)$ in Eq.~\eqref{eq:dos} for bilayer graphene without (a) and with (b) light-matter interaction as a function of energy and momentum. The path $(\bs{K}-0.7 \hat{\bs{y}})\rightarrow  \bs{K} \rightarrow (\bs{K}-0.7 \hat{\bs{K}})$ in momentum space is illustrated in Fig.~\ref{fig:fig3}(b) as red line. The coupling of the leads to layers is set to $\Gamma_{T,B} = 5 \times 10^{-3}\gamma_0$ and the grey solid line shows the doping level which is $\eps_D=-0.2\gamma_0$.   In (a) the quasienergies exhibit the quadratic dispersion for bilayer graphene  close to the $\bs{K}$-point and the two conduction/valence bands are separated by $\gamma_{\perp} = 0.15 \gamma_0$. 
	In (b), bilayer graphene is irradiated with frequency $\omega = 0.6\gamma_0$ and optical  coupling strength $\gamma_{\alpha} = 0.1\gamma_0$ in top and bottom lead.  The light-matter interaction induces gaps in the density of states. On the right y-axis we label the energies of the dynamical gaps with $\eps_n =  n \omega/2+\eps_D$ and $n= \pm 1, \dots, \pm 4$. These gaps correspond to the resonant absorption of $n$ photons between conductance and valence band.}   
	\label{fig:fig4}
\end{figure*}

\section{Results}
\label{sec:results}
In this section, we discuss the density of states and the transport properties of the bilayer heterostructure under light irradiation. We study two kinds of bilayer heterostructures: bilayer graphene and a graphene/hBN heterostructure. 
Our model comprises several parameters that we take from DFT calculations~\cite{giovannetti:2007,slawinska:2010} or estimate from experimental results~\cite{mciver:2019}.

\subsection{Discussion of parameters}
\label{sec:resultsparameters}
The stacking of the bilayer heterostructure has the lowest energy when one carbon atom of sublattice $a$ is over carbon (boron) atom and the carbon atom of sublattice $b$ is over the hollow site of the lower layer~\cite{giovannetti:2007,slawinska:2010} [Fig.~\ref{fig:fig3}(a)]. The lattice constants of graphene and hBN differ by a small distance justifying consideration of commensurate geometries~\cite{catellani:1987,castro:2009}. 
Different values for the in-plane next-nearest neighbor coupling $\gamma_{\parallel}$ and the out-of plane coupling $\gamma_{\perp}$ have been reported in bilayer graphene and graphene/hBN heterostructures~\cite{castro:2009,slawinska:2010,rozhkov:2016}. We set the in-plane and out-plane couling for both graphene and hBN to 
$\gamma^{}_{\parallel} = 2.6 \textrm{ eV}$ 
and
$\gamma^{}_{\perp} = 0.4 \textrm{ eV}$,
respectively. The energies of the $2p_z$ level in the carbon atom ($C$) are set to $\eps_{C,\alpha}^{} = 0$ and the energies of boron and nitride atoms are set to $\eps_{B,\alpha}^{} = 3.3\textrm{ eV}$ and $\eps_{N,\alpha}^{} =-\textrm{1.4 eV}$~\cite{slawinska:2010}. 
Since typical heterostructures are doped, we set a doping energy to $\eps_D=-0.2\gamma_0$ such that states both above and below the $\bs{K}$-point are occupied in an equilibrium state.

To get an estimate of the magnitude of the vector potential, we convert the vector potential $\bs{A}(t)$ to an electric field by $\bs{E}(t) = \partial \bs{A}(t)/\partial t $. In Ref.~\onlinecite{mciver:2019}, a graphene sample was illuminated with an electric field amplitude $E = 4.0\times 10^7\textrm{ V/m}$\cite{mciver:2019} and a photon energy $\hbar\omega \approx 191 \textrm{ meV}$. Introducing the interaction energy of next-nearest neighbor carbon atoms $\gamma_{\parallel}=2.6 \textrm{ eV}$, the photon energy corresponds to $ \hbar\omega \approx 0.07 \gamma_0$. The light-matter coupling strength then is of the order $\gamma_{\alpha} \approx 0.02 \gamma_0$ with the factor $K_\alpha \simeq 3.0 \text{ nm}^{-1} $ \cite{Malic:2011}.   For the purpose of presentation, we show in following sections the results for slightly larger electric fields and photon energies. For example, we assume an electric field amplitude of the order of $E= 0.1 \textrm{ V/\AA}$ and a photon energy $\hbar\omega \approx 1.7\textrm{ eV}$  corresponding to $\gamma_{\alpha} \approx 0.1 \gamma_0$ and $\hbar\omega \approx 0.6\gamma_0$.

\subsection{Spectral density of states}
\label{sec:resultsdos}

 \begin{figure*}[t!]			
	\includegraphics[width=0.47\linewidth,angle=0.]{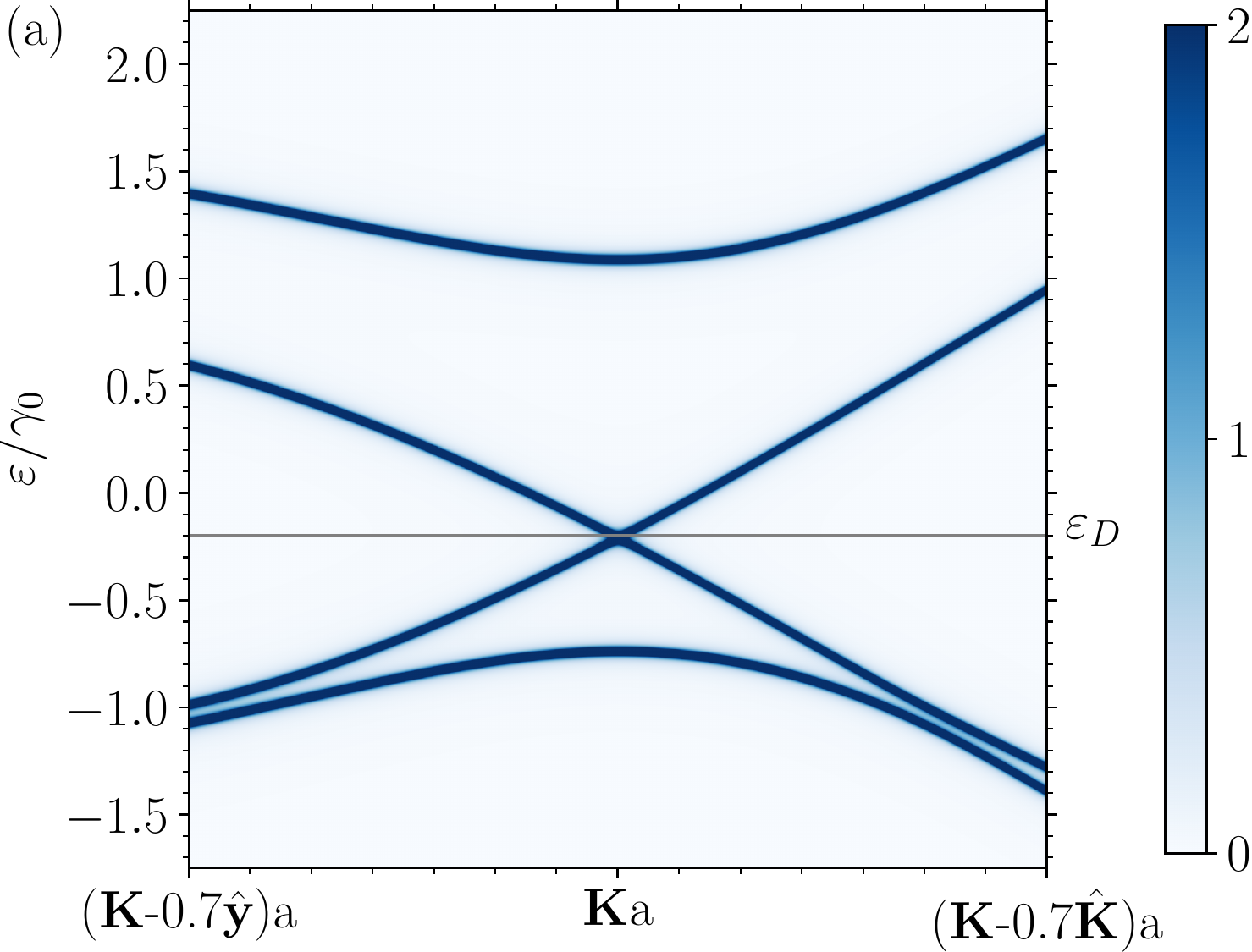} 
	\hspace{15pt}
	\raisebox{0mm}{\includegraphics[width=0.47\linewidth,angle=0.]{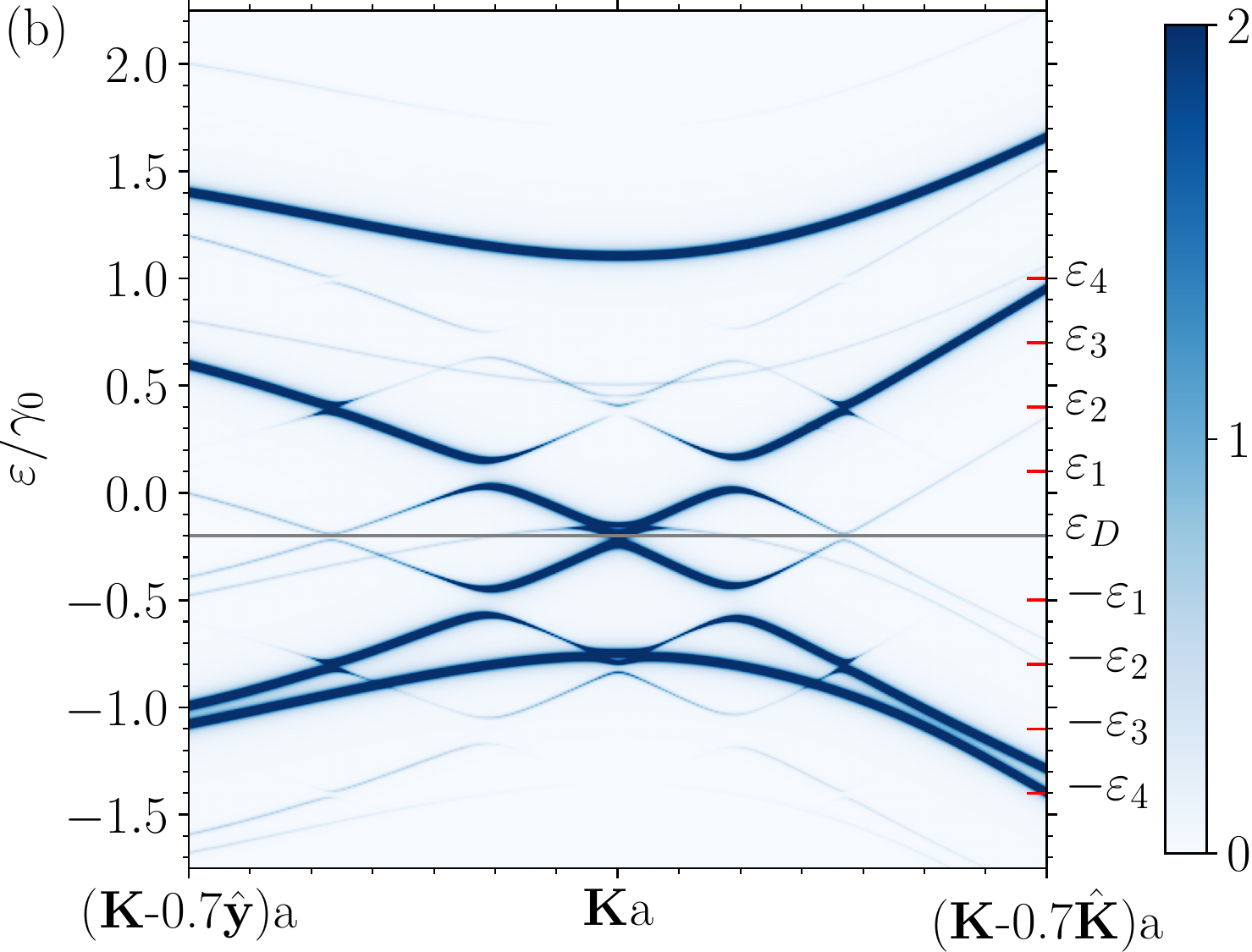}}
	\caption{Spectral density of states $\rho_{\bs{k}}(\eps)$ in Eq.~\eqref{eq:dos} for graphene/hBN heterostructure without (a) and with (b) light-matter interaction as a function of energy and momentum. The path $(\bs{K}-0.7 \hat{\bs{y}})\rightarrow  \bs{K} \rightarrow (\bs{K}-0.7 \hat{\bs{K}})$ in momentum space is illustrated in Fig.~\ref{fig:fig3}(b) as red line. The coupling of the leads to layers is set to $\Gamma_{T,B} = 5 \times 10^{-3}\gamma_0$ and the grey solid line shows the doping level which is $\eps_D=-0.2\gamma_0$.   In (a) the quasienergies are decomposed of a linear dispersion close to the $\bs{K}$-point which is mainly due to graphene, and two bands originating mainly from boron and nitrogen atoms. 
		In (b), the graphene/hBN heterostructure is irradiated with frequency $\omega = 0.6\gamma_0$ and optical coupling strength $\gamma_{\alpha} = 0.1\gamma_0$ in top and bottom lead.  The light-matter interaction induces gaps in the density of states. On the right y-axis we label the energies of the dynamical gaps with $\eps_n =  n \omega/2+\eps_D$ and $n= \pm 1, \dots, \pm 4$. These gaps correspond to the resonant absorption of $n$ photons between conductance and valence band.   }   
	\label{fig:fig5}
\end{figure*}

The energy and momentum-dependent spectral density of states $\rho_{\bs{k}}(\eps) $ of graphene under light irradiation is given by
\begin{equation}
\rho_{\bs{k}}(\eps) = -\frac{1}{4\pi}\mathrm{Im\,Tr\,}\hat{G}^{0,R}_{\bs{k}}(\eps) \, ,
\label{eq:dos}
\end{equation}
with the retarded Floquet Green's function $\hat{G}^{0,R}_{\bs{k}}(\eps)$ at $n=0$ photons~\cite{calvo:2013}. Figure~\ref{fig:fig4} shows the spectral density of states of bilayer graphene and graphene/hBN close to the $\bs{K}$-point. The path along which we show the density of states is illustrated in Fig.~\ref{fig:fig3}(b) as red line. 
It stretches from
$
(\bs{K}-0.7 \hat{\bs{y}})\rightarrow  \bs{K} \rightarrow (\bs{K}-0.7 \hat{\bs{K}})
$ 
in the first Brillouin zone with the unit vectors $\hat{\bs{y}}$ and $\hat{\bs{K}}$ in $\bs{y}$ and $\bs{K}$-direction, respectively. 

Comparing Fig.~\ref{fig:fig4} (a) and (b), the light-matter interaction induces sidebands that are shifted by $n\hbar\omega$ from the original band without light-matter interaction. For the parameters in Fig.~\ref{fig:fig4}(b), only sidebands with $\pm 1$ shifted from the bare quasienergy spectra are visible in the spectral density of states. The effect of light on the quasienergy spectrum decreases with the number of sidebands and we restrict the calculation to $n=4$ sidebands. We also tested that higher orders will not change the results.

Besides the introduction of sidebands, the interaction of light with the bilayer heterostructures causes opening of gaps in the quasienergy spectra. In literature, these gaps have been separated into gaps that occur at doping energy $\eps_D$ and at energies away from $\eps_D$. For example, it has been discussed that the gap for undoped graphene at zero energy leads to a Hall current without a magnetic field for circularly polarized light~\cite{oka:2009}. 
Gaps away from the doping energy are also called dynamical gaps~\cite{syzranov:2008} and occur at energies $\eps_n =  n \omega/2+\eps_D$ with the number sidebands $n= \pm 1, \dots, \pm 4$. These gaps correspond to a resonant absorption/emission of $n$ photons with energy $\hbar \omega$ between the conduction and valence band. Here, we focus the discussion on the dynamical gaps since the dominant signatures of the light-matter interaction in the differential conductance (see Sec.~\ref{sec:resultscond}) are related to those gaps.

The energies $\eps_n$ of the dynamical gaps are indicated on the right axis in Fig.~\ref{fig:fig4}(b). 
The largest gap opens at a resonant absorption of a single photon and occurs at energies $\eps_{\pm 1} =  \pm \omega/2+\eps_D$. For these energies, the quasi-energy spectra becomes fully gapped close to the $\bs{K}$-point.
However, further away from the $\bs{K}$-point copies of the bare spectra of the bilayer heterostructure intersect energies at $\eps_1$. These states at energy $\eps_1$ will be responsible for a finite differential conductance at voltages $eV\simeq \eps_1$ as we will discuss in Sec.~\ref{sec:resultscond}.

Figure~\ref{fig:fig5} shows the spectral density of states for a graphene/hBN heterostructure. Without the light-matter interaction, the band structure consists of linear dispersion due to the graphene layer two bands due to hBN. The valence band with $\bs{K}$-point energy of $\eps=-0.7\gamma_0$ is mainly constituted by the nitrogen
sublattice while the conduction band with $\bs{K}$-point energy of $\eps=1.1\gamma_0$  originates from boron atoms. Under light-irradiation gaps open at energies $\eps_n$ similar to the case of bilayer graphene. The bands which are mainly due to nitrogen and boron show a weaker dependence on momenta near the $\bs{K}$-point in comparison to the linear dispersion of graphene. Such behavior implies a large density of states at energies $\eps =1.1\gamma_0$ and $\eps=-0.7\gamma_0$ suggesting that coupling to light induces additional features in the differential conductance as discussed in Sec.~\ref{sec:resultscond}. Such features also appear in bilayer graphene at larger energies.

It is interesting to note that the size of the gaps in both Fig.~\ref{fig:fig4} and \ref{fig:fig5} depend on direction from which the $\bs{K}$-point is approached. The different gaps sizes are due to the momentum-dependence of the optical matrix elements ${M}_{\bs{k},\alpha}$ in Eq.~\eqref{eq:Hlightmatter}. As an example,  when changing the momentum from $(\bs{K}-0.7 \hat{\bs{y}})$ to the $ \bs{K} $-point the gap at energy $\eps_1$ in Fig.~\ref{fig:fig4}(b) is slightly smaller compared to the gap at energy $\eps_1$ and momenta between $ \bs{K} $ and $ (\bs{K}-0.7 \hat{\bs{K}}) $.

\subsection{Differential conductance}
\label{sec:resultscond}
In this section we discuss the differential conductance of Eq.~\eqref{eq:diffcond} 
for both bilayer graphene and graphene/hBN. We can separate signatures of the light-matter interaction into processes which are related to the opening of gaps as discussed in Sec.~\ref{sec:resultsdos}, and into processes that are related to the tunneling into and out of a Van~Hove singularity. The first kind of processes leads to a suppression of the conductance, the second kind to an enhancement. We discuss the behavior of the differential conductance at zero temperature and assume that the voltage is applied to the top lead. 

\begin{figure}[t!]			
	
	\begin{center}
		\includegraphics[width=0.95\linewidth,angle=0.]{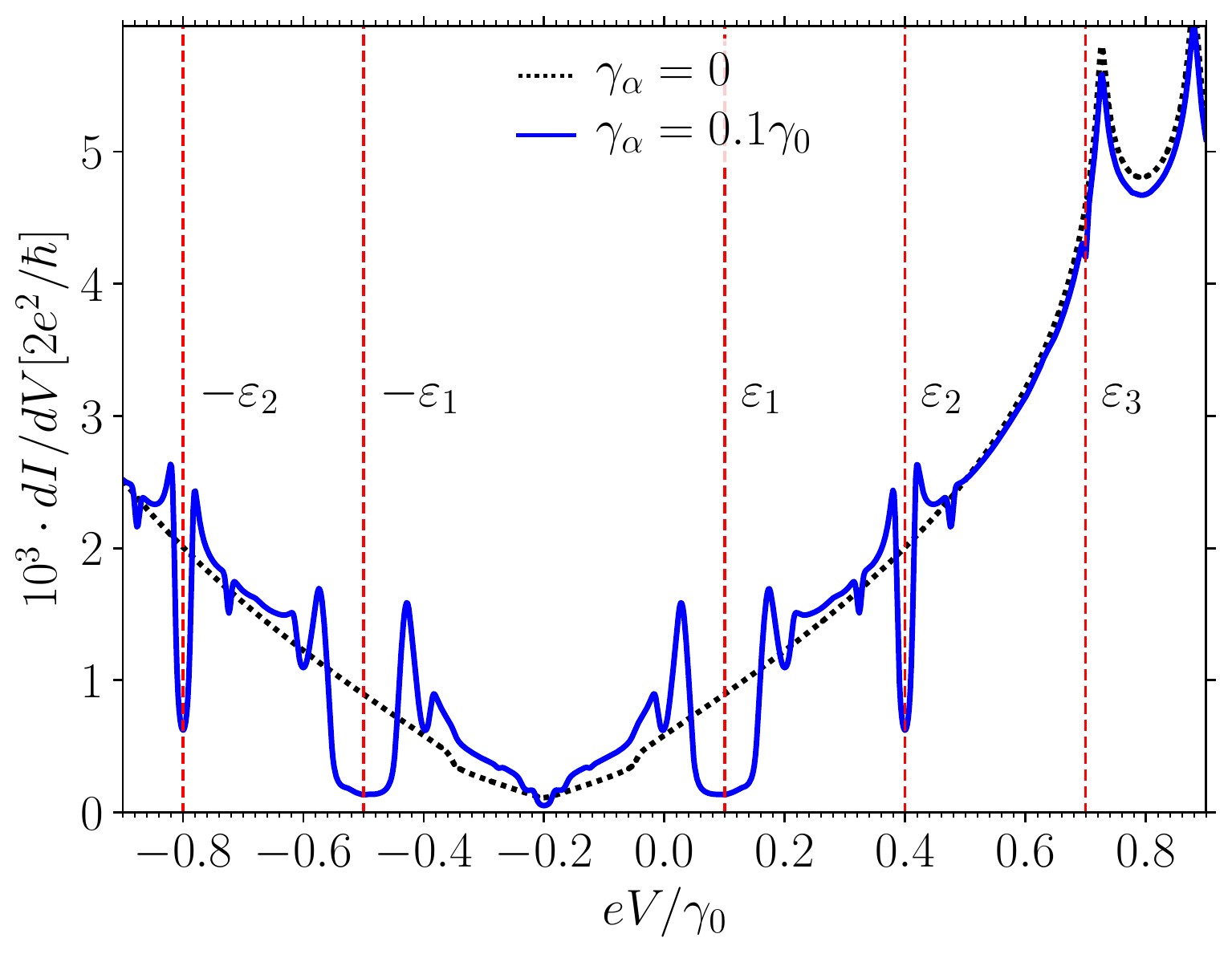} 
	\end{center}
	\caption{Differential conductance for bilayer graphene. The parameters are the same as in Fig.~\ref{fig:fig4}. The voltage is applied on the top lead and the dashed black line shows the conductance without light. Gaps in the quasienergy spectra suppress the current at voltages $eV = \eps_n$.  The parameters are $\omega = 0.6 \gamma_0$, $T=0$, $n=4$, $\Gamma_{T,B} = 5 \times 10^{-3}\gamma_0$. For the parameters shown in the figure gaps occurs at voltages $\eps_1$, $\eps_2$ and $\eps_3$ indicated as red dashed line and also shown in Fig.~\ref{fig:fig4}(b).}   
	\label{fig:fig6}
\end{figure}

Figure~\ref{fig:fig6} shows the differential conductance of bilayer graphene when light is turned on (blue line) and off (black dashed line), respectively. As expected the conductance is suppressed near voltages corresponding to energies $\eps_n$ of the gaps in the quasi-energy spectra of bilayer graphene.
The energies $\eps_n$ are shown in Fig.~\ref{fig:fig4}(b) and Fig.~\ref{fig:fig6} as dashed red lines. For the voltage range of Fig.~\ref{fig:fig6} gaps occur at $\pm\eps_1$, $\pm\eps_2$ and $\eps_3$. The width of the gaps decreases with increasing number $n$ of sidebands. 
Although in Fig.~\ref{fig:fig4}(b) a gap with width $\simeq 0.1 \gamma_0$ opens near energy  $\eps_1$, the conductance has a small but finite value. As we discussed in Sec.~\ref{sec:resultsdos}, the gap at $\eps_1$ opens close to the $\bs{K}$-point. However, away from the $\bs{K}$-point along the paths $(\bs{K}-0.7 \hat{\bs{y}})\rightarrow  \bs{K}$ and $\bs{K} \rightarrow (\bs{K}-0.7 \hat{\bs{K}})$ copies of the conductance band intersect the gap at energy $\eps_1$ [see Fig.~\ref{fig:fig4}(b)]. The copies of the conductance band are the reason for the finite conductance close to voltages $eV\simeq\eps_1$. The same arguments holds for $eV\simeq-\eps_1$.

\begin{figure*}[t!]			
	
	\begin{center}
		\includegraphics[width=0.46\linewidth,angle=0.]{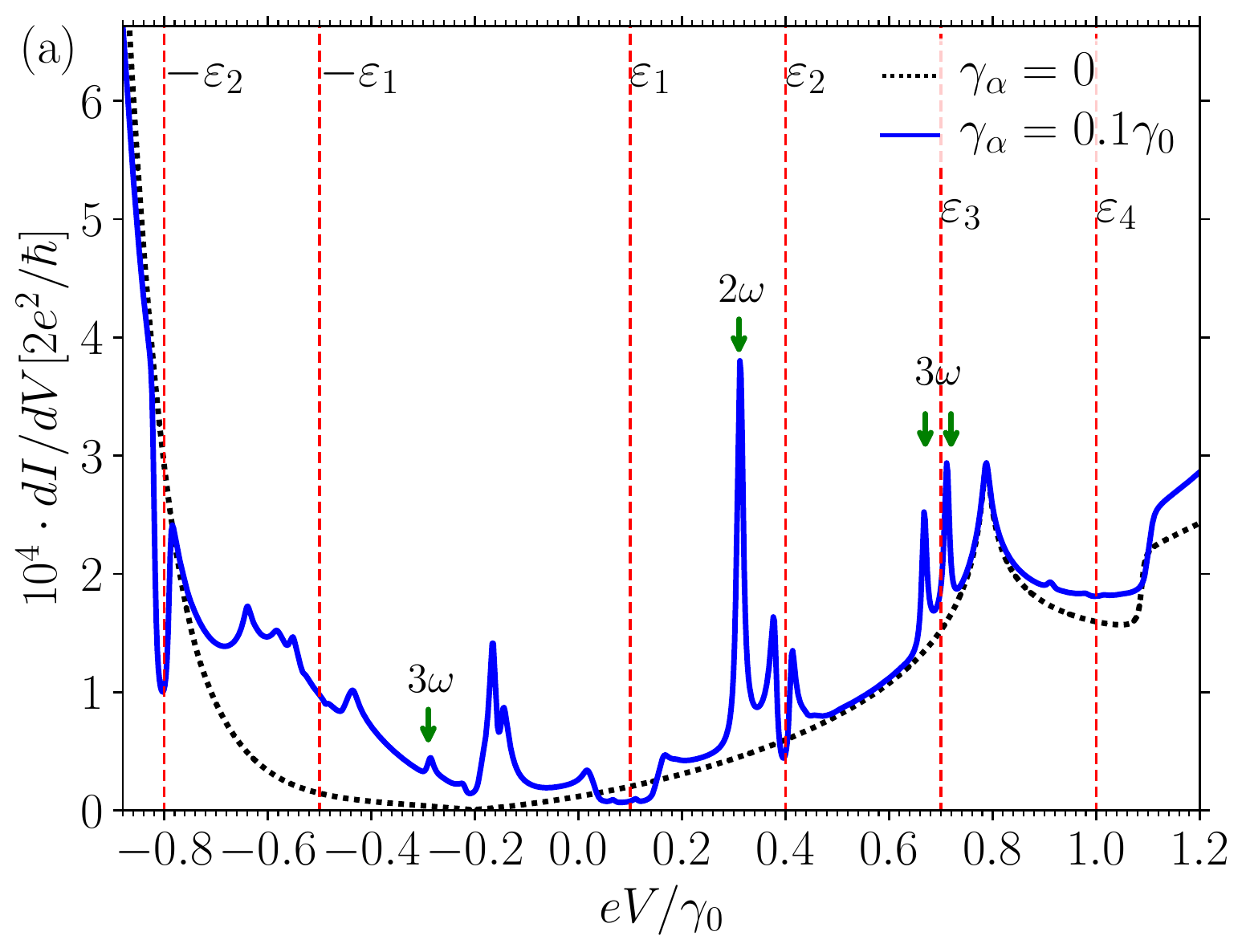} 
		\hspace{15pt}
		\raisebox{0mm}{\includegraphics[width=0.483\linewidth,angle=0.]{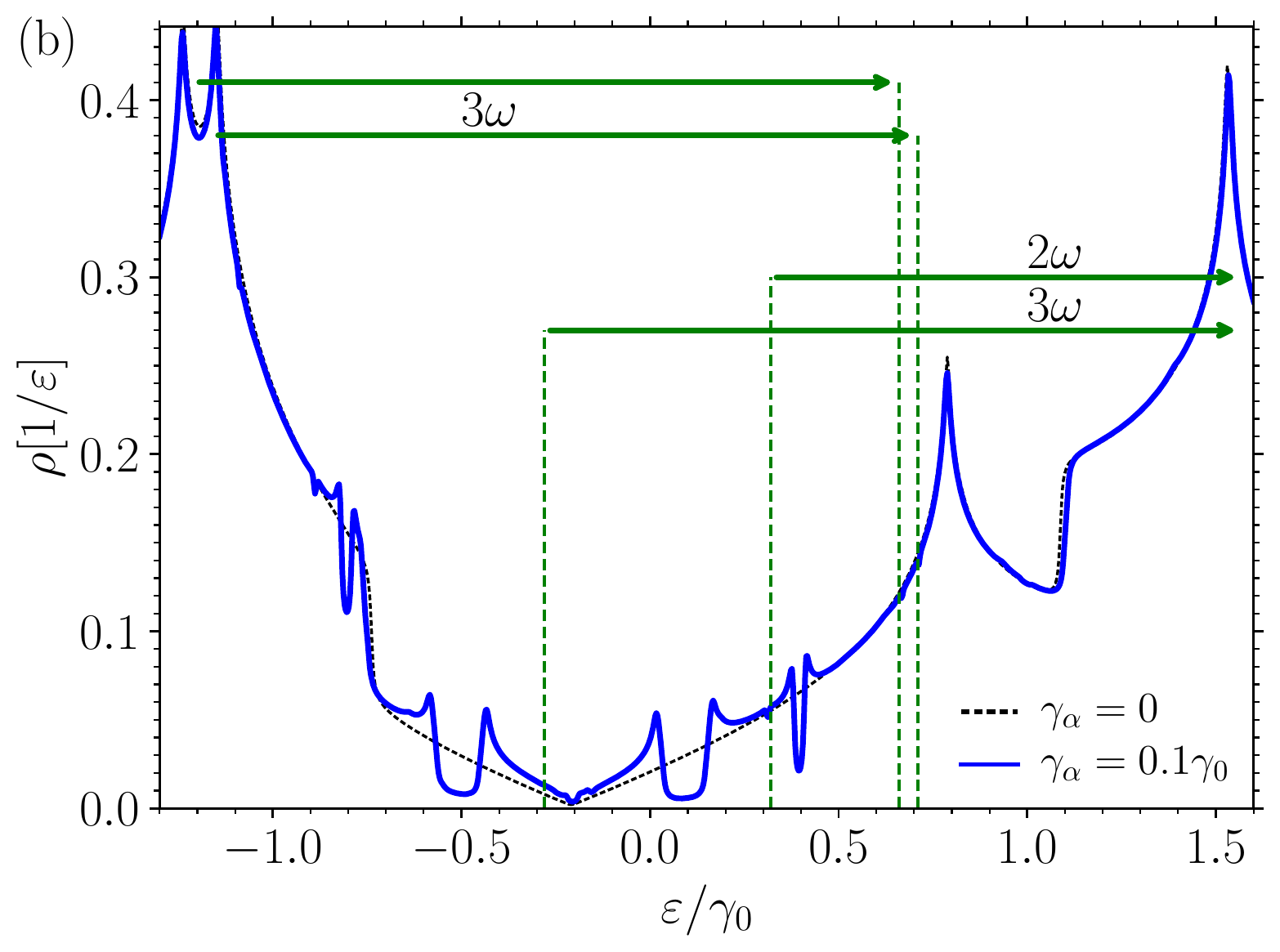} }
	\end{center}
	\caption{Differential conductance (a) and density of states (b) for a graphene/hBN heterostructure. The parameters in (a) and (b) are same as in Fig.~\ref{fig:fig5}. The voltage is applied at the top lead and the dashed black line shows the conductance without light. Gaps in the quasienergy spectra suppress the current at voltages $eV = \eps_n$.  At certain voltages (indicated as green vertical arrows in (a)) an enhancement of the conductance occurs which is related to the tunneling in and out-of a van-Hove singularity. The processes giving rise to the peaks indicated with green arrow in (a) are sketched in (b) as green arrows. As example, the peak at $eV\simeq 0.31\gamma_0$ originated from two-photon absorption and tunneling into the Van~Hove singularity at $eV\simeq 1.51\gamma_0$. The parameters are $\omega = 0.6 \gamma_0$, $T=0$, $n=4$, $\Gamma_{T,B} = 5 \times 10^{-3}\gamma_0$. }   
	\label{fig:fig7}
\end{figure*}

In addition to the suppression of the conductance at voltages $\eps_1$, $\eps_2$ and $\eps_3$, smaller modifications of the conductance occur at other voltages. For example, close to the voltage $eV\simeq \eps_2$  two small dips appear at slightly larger and smaller voltages than $eV\simeq\eps_2$. These dips can be understood by considering the quasienergy spectra in Fig.~\ref{fig:fig4}(b) close to energy $\eps_2$. Slightly above and below the energy $\eps_2$  in Fig.~\ref{fig:fig4}(b), there are additional avoided crossing stemming from the two conductance bands of bilayer graphene. These gaps are responsible for the small dips close to $\eps_2$. 

We now turn to the differential conductance in graphene/hBN heterostructures which is shown in Fig.~\ref{fig:fig7}(a). To explain some features in the differential conductance, it is instructive to discuss the total density of states in Fig.~\ref{fig:fig7}(b). The total density of states is obtained from
$
\rho(\eps) = (1/N_{\bs{k}}) \sum_{\bs{k}}\rho_{\bs{k}}(\eps) 
$
with the total number of $\bs{k}$-point $N_{\bs{k}}$ in the Brillouin zone.

Similar to Fig.~\ref{fig:fig6} the differential conductance is suppressed at energies $\eps_n$ with the width decreasing with an increasing number of sidebands. Additional features originate from tunneling into or out of a Van~Hove singularity. The Van~Hove singularities provide a large number of states resulting in an enhancement of the differential conductance.

Similar to the gaps that occur due to absorption/emission of $n$-photons, the signatures due to tunneling in or out of a Van~Hove singularity can be divided into $n$-photon processes. The most dominant feature in Fig.~\ref{fig:fig7}(a) occurs due to a two-photon process and is sketched in Fig.~\ref{fig:fig7}(a) as green arrow. At $eV\simeq 0.31 \gamma_0$,  electrons can tunnel to the Van~Hove singularity at energies $eV\simeq 0.31 \gamma_0 + 2 \omega$ by absorption of two-photons. In addition to the two-photon tunneling process, an electron can also be absorbed and tunneling to the same Van~Hove singularity via a three-photon tunneling process. In this case, the differential conductance shows a small peak at negative voltages $eV\simeq -0.31 \gamma_0$. 

A similar process occurs at voltages $eV\simeq 0.7 \gamma_0$ near the gap opening at energy $\eps_3$. Although the opening of a gap implies the suppression of the differential conductance, near voltages $eV\simeq \eps_3$ the dominant process is related to tunneling out of a Van~Hove singularity resulting in an enhancement of the conductance. The corresponding three-photon process is sketched in Fig.~\ref{fig:fig7}(b). In this case the Van~Hove singularities at $eV \simeq -1.2 \gamma_0$ which are separated by $\simeq 0.1\gamma_0$ provide a large density of states resulting in two peaks at $eV\simeq 0.7 \gamma_0$ via a three-photon tunneling process.

Further interesting features of the light-matter interaction occur at negative voltages in the range $-0.8 \gamma_0\lesssim eV \lesssim -0.2 \gamma_0$. In this range, the differential conductance without light is much smaller than the differential conductance under light-irradiation. This can be understood by the spectral density in Fig.~\ref{fig:fig5}. At $eV = -0.2 \gamma_0$, the two upper bands in Fig.~\ref{fig:fig5}(a) are empty while the two lower bands are occupied. If the voltages is reduces to $eV \lesssim -0.2 \gamma_0$, a further band is emptied close to the \textbf{K}-point. Under light-irradiation, electrons in the lowest energy band can then be absorbed and tunnel into the empty band. Such processes can occur for $-0.8 \gamma_0\lesssim eV \lesssim -0.2 \gamma_0$. For voltages $eV<-0.8 \gamma_0$, the lowest-lying energy band is also emptied and electrons can't be absorbed. Hence, in the range $-0.8 \gamma_0\lesssim eV \lesssim -0.2 \gamma_0$, the differential conductance is offset from the conductance without light.

For the parameter regime that we discussed in this paper, tunneling in and out of a Van~Hove singularity is more dominant in graphene/hBN heterostructure than in bilayer graphene. The reason for this is that the conductance of bilayer graphene is an order of magnitude smaller than the conductance of the graphene/hBN heterostructure. Since the light-matter coupling parameters are the same in both cases, the light-matter interaction is more pronounced in the graphene/hBN heterostructure.
Another reason is that the opening of gaps suppresses the process of tunneling into a Van~Hove singularity. For example, a one-photon process corresponding to the tunneling into the Van~Hove singularity at $eV\simeq 0.72\gamma_0$ in Fig.~\ref{fig:fig6} is suppressed by the gap near voltages $eV\simeq \eps_1$. However, a tiny peak appears at $eV\simeq 0.28\gamma_0$ corresponding to the absorption of a single photon and the tunneling into the Van~Hove singularity at $eV\simeq 0.88\gamma_0$. The peak $eV\simeq 0.88\gamma_0$ is barely visible due to the two-bands of bilayer graphene giving rise to a larger conductance without light compared to graphene/hBN heterostructures.

%
%
%
%
\section{Conclusion}
\label{sec:conclusions}
We studied the transport properties of vertical heterostructures under light irradiation in Landauer-type configuration. The vertical heterostructure is connected to leads which are assumed to be in equilibrium if the central region is irradiated with light. As particular examples for a central region, we considered bilayer graphene and hexagonal boron-nitride heterostructures. In such contact, we calculated the differential conductance by using the nonequilibrium Green's function formalism.

The external light-field induces sidebands of the original band structure and opening of gaps which correspond to a resonant absorption or emission $n$-photons between the conductance and valence band. 
These gaps occur at energies $\eps = n \omega/2 + \eps_D$ with the frequency $\omega$ of the photons and the doping energy $\eps_D$. In transport properties, the gaps are manifested in a strong suppression of the conductance. Although gaps open close to the $\bs{K}$-point, at momenta further away from the $\bs{K}$-point sidebands intersect the gaps. These sidebands give rise to a finite conductance even when a gap occurs close to the $\bs{K}$-point. 
Besides the suppression, light-matter interaction can also lead to an enhancement in the differential conductance. The enhancement of the differential conductance is related to the tunneling of photo-excited electrons into or out of van Hove singularities of the vertical heterostructure. 

Processes originating from tunneling into or out of a Van~Hove singularity prevail in graphene/hBN heterostructures compared to bilayer graphene due to the smaller conductance in graphene/hBN heterostructures. In general, suppression of conductance due to opening of gaps compete with processes due to tunneling into or out of Van~Hove singularities. For example, the absorption of two-photon and the accompanied tunneling into the Van~Hove singularity of graphene/hBN heterostructures is larger the suppression due to opening of gaps.

\acknowledgments 
This work was supported by the Swedish Foundation for Strategic Research (GMT14-0077). We further acknowledge the funding provided by 2D TECH VINNOVA competence Center (Ref. 2019-00068).


\bibliographystyle{apsrev-titles}
\bibliography{references}

\end{document}